\documentclass[twocolumn,aps,prb]{revtex4}
\usepackage{graphicx}

\begin{document}
\title{X-ray diffraction peak profiles from threading dislocations in
       GaN epitaxial films}
\author{V. M. Kaganer}
\author{O. Brandt}
\author{A. Trampert}
\author{K. H. Ploog}
\affiliation{Paul-Drude-Institut f\"{u}r Festk\"{o}rperelektronik,
Hausvogteiplatz 5--7, D-10117 Berlin, Germany}

\begin{abstract}
We analyze the lineshape of x-ray diffraction profiles of GaN epitaxial
layers with large densities of randomly distributed threading
dislocations. The peaks are Gaussian only in the central, most intense
part of the peak, while the tails obey a power law. The $q^{-3}$ decay
typical for random dislocations is observed in double-crystal rocking
curves. The entire profile is well fitted by a restricted random
dislocation distribution. The densities of both edge and screw threading
dislocations and the ranges of dislocation correlations are obtained.
\end{abstract}

\date{\today}


\maketitle

\section{Introduction}

GaN epitaxial layers grown on different substrates (e.g.,
Al$_{2}$O$_{3}$, SiC, or Si) possess very large densities of threading
dislocations which cross the layer along its normal, from the
layer-substrate interface to the surface.\cite{gilbook} The threading
dislocation density depends only marginally on the substrate material
(and hence on the misfit between the substrate and the layer) but rather
on the growth technique and conditions. For (0001) oriented layers of
wurtzite GaN, the overwhelming majority of dislocations are of edge type
with Burgers vectors $\mathbf{b}=\frac{1}{3} \left\langle
11\overline{2}0\right\rangle $. Their
density\cite
{srikant4286jap,metzger1013pma,heinke2145apl,sun02,chierchia03} is
typically $10^{8}-10^{10}$~cm$^{-2}$. The density of screw dislocations
with Burgers vector $\mathbf{b}=[0001]$ is one to two orders of magnitude
lower than the density of edge dislocations.

\begin{figure}[tbp]
\noindent
\includegraphics[width=\columnwidth]{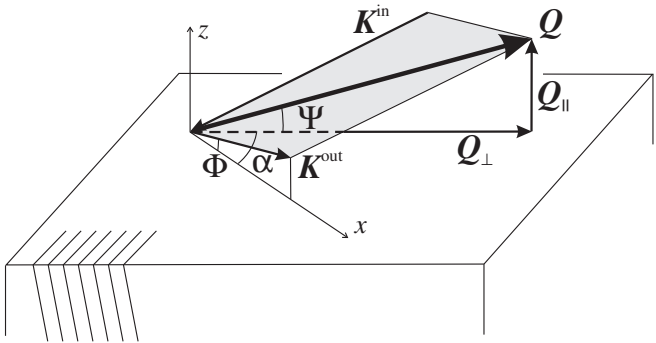} \caption{Sketch of skew
geometry x-ray diffraction. The lattice planes of the actual reflection
are depicted in the left bottom part of the figure. The incident wave
$\mathbf{K}^{\rm in}$ and the diffracted wave $\mathbf{K}^{\rm out}$ make
the same angle $\Phi$ to the sample surface. The scattering vector
$\mathbf{Q}$ makes an angle $\Psi$ to the sample surface.}
\label{skew}
\end{figure}

The dislocation density can be measured directly by plan-view
transmission electron microscopy (TEM). The actual virtue of TEM is not
the accurate determination of the dislocation density, but rather the
possibility to determine the type of the dislocation distribution (e.~g.,
random vs.\ granular/columnar structure). It is of limited statistical
significance considering the small area covered by TEM micrographs.
Alternatively, x-ray diffraction can be used to detect the lattice
distortions due to the presence of dislocations. In principle, both the
dislocation distribution and the actual dislocation density may be
obtained from x-ray diffraction profiles.

The impact of screw and edge threading dislocations on the width of the
x-ray reflections in the limiting cases of lattice planes parallel and
perpendicular to the surface is commonly referred to as tilt and twist,
respectively.\cite
{srikant4286jap,metzger1013pma,heinke2145apl,sun02,chierchia03}
This designation stems from the model of misoriented
blocks\cite{chierchia03} which is \emph{not} appropriate for strains
caused by randomly distributed dislocations. In this case, the
description in terms of mean-squared distortions\cite{ratnikov00} is
actually more adequate. Moreover, all previous experimental studies
only use the full width at half-maximum (FWHM) of different
reflections as a measure of the dislocation density.

The symmetric Bragg reflections from GaN layers are comparatively narrow,
since they are not influenced by the presence of edge dislocations. Edge
dislocations produce distortions within lattice planes parallel to the
surface but do not disturb positions of these planes along the layer
normal. The highest sensitivity to edge dislocations is obtained by
diffraction from lattice planes perpendicular to the surface. This
diffraction geometry requires grazing incidence illumination and is thus
commonly performed at a synchrotron.\cite{munkholm2972apl,yang4240jap} An
alternative geometry that can easily be realized in the laboratory is the
skew geometry,\cite{srikant4286jap,sun02} which is a quasi-symmetric (the
incident and the diffracted waves have the same angles to the surface)
non-coplanar (the surface normal does not lie in the plane defined by the
incident and the diffracted waves) geometry, as shown in Fig.\
\ref{skew}. By measuring different reflections with increasing lattice
plane inclination, one can extrapolate to lattice planes perpendicular to
the surface.\cite{srikant4286jap,sun02} A single reflection with a large
inclination can be regarded as a figure of merit.\cite{heinke2145apl} A
four-circle diffractometer\cite{busing457ac} is required for skew
geometry measurements, since the sample has to be tilted with respect to
its surface normal. Coplanar asymmetric reflections on a three-circle
diffractometer are much less sensitive to edge dislocations since they
only partially touch the lattice distortions parallel to the surface
plane.\cite{sun02}

The FWHM of the diffraction peak depends not only on the dislocation
density, but also on the correlations between dislocations. Furthermore,
it depends on the mutual orientations of the scattering vector,
dislocation line direction, and the Burgers vector direction, which have
not been taken into consideration in previous diffraction studies of GaN
layers. In the present paper, we analyze the entire lineshape of the
diffraction profiles, and particularly their tails. These tails are due
to scattering in the close vicinity of the dislocation lines and are not
influenced by the correlations between dislocations. They follow
universal power laws and can be used to determine the dislocation
density. We fit the entire diffraction peak profile to the numerical
Fourier transform of the pair correlation function and simultaneously
obtain dislocation densities and the range of dislocation correlations.

\section{Background}

X-ray diffraction is a well-established technique to analyze crystal
lattice imperfections.\cite{warren:book69,krivoglaz:book} The
conventional and widely used methods (also in GaN
studies\cite{chierchia03}) are based on a comparison of the diffraction
peak widths of different reflections.\cite
{WilliamsonHall53,hordon61,ayers94} They aim to separate two
contributions to the peak broadening, the finite size of the crystalline
domains (grains) in the sample and the non-uniform strain within each
domain owing to lattice defects. The strain broadening of a diffraction
peak is proportional to the length of the scattering vector, while the
size effect does not depend on it. Thus, comparing the peak widths of
reflections of successive orders in the so-called Williamson-Hall plot,
one can separate both contributions. The separation method assumes that
all peaks are Gaussian,\cite{hordon61,ayers94} or all
Lorentzian.\cite{WilliamsonHall53} This is certainly not true, and the
methods based on the peak width give only rough, albeit instructive,
estimates of the crystal perfection. A recent development
\cite{ungar96,ungar99,ungar01} includes corrections for given
orientations of the dislocation lines and Burgers vectors with respect to
the scattering vector.

The next milestone was the Fourier analysis of the diffraction peak
profile proposed by Warren and
Averbach.\cite{WarrenAverbach50,WarrenAverbach52} The starting point of
their analysis is the average over grain orientations in powder
diffraction. The corresponding integration of the scattered intensity in
reciprocal space is equivalent to a one-dimensional cut of the
correlation function in real space. We note that the powder diffraction
literature does not use the term correlation function but refers to the
Fourier coefficients of the intensity. We use the solid-state physics
terminology where this quantity is commonly called the pair correlation
function. In crystallography, the same function is referred to as
Patterson function.

The Ansatz of the Warren-Averbach analysis is the assumption that the
correlation function is a product of two independent terms describing
the size and the strain effects, respectively. Furthermore, it is assumed
that the finite sizes of the grains give rise to a Lorentzian peak
(exponential correlation function) while non-uniform strain is described
by a Gaussian function. With these assumptions, the size and strain
effects are separated by differentiation of the Fourier-transformed peak
profile. In many cases, the latter step is not sufficiently accurate,
since it has to rely on a few Fourier coefficients obtained from noisy
experimental data. Balzar\cite{balzar92,balzar93} suggested to avoid this
difficulty by directly fitting the peak profile to a convolution of a
Lorentzian and a Gaussian, which is the Voigt function. The experimental
peak profiles are not always described by the Voigt function and various
other analytical functions were suggested on a purely phenomenological
basis.\cite {young82,keijser83}

The assumption that non-uniform strain gives rise to a Gaussian
correlation function looks plausible, taking into account the stochastic
nature of this strain originating from randomly distributed lattice
defects. However, Krivoglaz and Ryboshapka\cite
{krivoglaz63,krivoglaz:book} have shown that this is not true for random
dislocations, which are the most common and most important source of
strain. Rather, they found that the slow ($\propto r^{-1}$) decay of the
strain with the distance $r$ from the dislocation line gives rise to a
correlation function
\begin{equation}
G(x)=\exp (-C\rho x^{2}\ln \frac{L}{\xi x}).  \label{eq1}
\end{equation}
Here $C\sim 1$ is a dimensionless factor depending on the orientations of
the dislocation line and Burgers vector with respect to the $x$-direction
(an arbitrary direction along which the correlations are measured). Its
dependence on the scattering vector $\mathbf{Q}$ and the Burgers vector
$\mathbf{b}$ is given by $C\propto (\mathbf{Q}\cdot \mathbf{b})^{2}$. The
dislocation density $\rho $ is defined as a total length of the
dislocation lines per unit volume. For straight dislocations, $\rho $ is
equal to the number of dislocations crossing the unit area of a plane
perpendicular to the dislocation lines. $\xi \sim 1$ is another
dimensionless factor, for uncorrelated dislocations $\xi =\left|
\mathbf{Q}\cdot \mathbf{b} \right| /2\pi $.

For completely random and uncorrelated dislocations, $L$ is the crystal
size, so that the diffraction peak width of an infinite crystal with
random uncorrelated dislocations tends to infinity. Wilkens\cite
{wilkens69,wilkens70nbs,wilkens70pss} pointed out that this divergence
has the same origin as the divergence of elastic energy of a crystal with
dislocations: the elastic energy is proportional to $\mu b^{2}\ln L/a$,
where $\mu $ is the shear modulus, $b$ is the length of the Burgers
vector, and $a$ is the lattice spacing. He suggested that the system can
drastically reduce elastic energy by a very minor rearrangement in the
dislocation ensemble: the positions of the dislocations remain random but
their Burgers vectors are correlated, so that the total Burgers vector in
a region exceeding some characteristic scale is zero. Then, the
dislocations screen each other and the elastic energy, as well as the
diffraction peak width, remain finite. Wilkens introduced a ``restricted
random distribution'' of dislocations by subdividing the crystal into
cells, such that the total Burgers vector in each cell is zero. He
found\cite {wilkens69,wilkens70nbs,wilkens70pss} that the functional form
of Eq.\ (\ref {eq1}) does not change but $L$ should be understood as a
finite size of the cells. Krivoglaz \emph{et
al.}\cite{krivoglaz83,krivoglaz:book} showed that the same result is
valid for a broad class of correlated dislocation distributions with
screening.

We note also the two-dimensional crystal as a limiting case of
dislocation screening. Here the elastic energy of dislocations is to be
compared with the entropic term $-TS$, where $T$ is temperature and the
entropy $S=\ln (L/a)^{2}$ is the number of lattice sites where the
dislocation can be placed. Both elastic energy and entropy terms are
proportional to $\ln L$. As a result, above some temperature $T_{m}$,
dislocations are generated by unbinding of thermally excited dislocation
pairs, giving rise to the dislocation mediated
melting.\cite{KosterlitzThouless72,KosterlitzThouless73} The calculation
of the correlation function in this highly correlated dislocation
system\cite{kaganer94prl} shows that the logarithmic term in Eq.\
(\ref{eq1}) vanishes.

Fourier transformation of the correlation function (\ref{eq1}) yields a
Gaussian shape only in the central part of the peak. The range of the
Gaussian peak shape is given by a dimensionless factor
\begin{equation}
M=L\sqrt{\rho }
\label{eq1a}
\end{equation}
and increases when $M$ is increased. The intensity distribution notably
deviates from the Gaussian shape at the tails of the diffraction peak.

The tails of the diffraction peak due to dislocations follow a universal
law $I(q)\propto q^{-3}$, which can be obtained from Fourier
transformation of Eq.\ (\ref{eq1}).\cite{groma98} The $q^{-3}$ law is due
to the fact that at large $q$, the scattering takes place in the strained
regions close to dislocation lines, where the lattice is so strongly
misoriented that the Bragg law is locally satisfied for the wave vector
$q$. Calculation of the volume of these regions give the $q^{-3}$
dependence.\cite {wilkens63,krivoglaz:book,groma98} Groma and
co-workers\cite {groma98,groma00,borbely01} developed methods for the
peak profile analysis based on a calculation of the restricted moments of
$I(q)$. In particular, the second-order restricted moment $v(q)=
\int_{-q}^{q}q^{2}I(q)dq $ is proportional to $\ln q$, which they used to
determine the mean dislocation density. Higher-order moments describe
fluctuations of the dislocation density.

GaN epitaxial film comprise a well-defined system where threading
dislocations are aligned perpendicular to the surface plane. The film is
oriented, contrary to a powder. However, the x-ray diffraction
measurements performed with an open detector give rise to an average very
similar to the powder average. The intensity is integrated over
directions of the outgoing beam, instead of the integration over
directions of the diffraction vector. The integrations in reciprocal
space gives rise to cuts in real space, which are different for the two
cases under consideration. The coordinate $x$ in Eq.\ (\ref{eq1}) runs
along the direction of the outgoing wave in case of the oriented sample
with open detector and in the direction of the diffraction vector for the
case of powder diffraction. This fact introduces a geometrical correction
in the orientational factor $C$. The skew diffraction geometry used for
the measurements gives rise to further corrections, which are calculated
below.

Our approach consists of a direct fit of the measured intensities by the
numerical Fourier transformation of the correlation function (\ref{eq1}),
thus avoiding any transformation of the experimental data. We expect that
such a calculation is less influenced by the noise in the experimental
data and is more reliable. As shown below, we find good agreement between
measured and calculated peak profiles. From the fits, we reliably obtain
both the dislocation densities and the correlation range in the
restricted random dislocation distribution.

\section{Theory}
\label{sec:theory}

The intensity of x-ray scattering from a crystal disturbed by strain
fields of lattice defects can be represented as a Fourier transform
\begin{equation}
I(\mathbf{q})=\int G(\mathbf{r})\exp (i\mathbf{q}\cdot \mathbf{r})
d\mathbf{r}
\label{eq2}
\end{equation}
of the pair correlation function
\begin{equation}
G(\mathbf{r})=\left\langle \exp \{i\mathbf{Q}\cdot [\mathbf{U}
(\mathbf{r})- \mathbf{U}(0)]\}\right\rangle .
\label{eq3}
\end{equation}
Here $\mathbf{Q}=\mathbf{K}^{\mathrm{out}}-\mathbf{K}^{\mathrm{in}}$ is
the scattering vector ($\mathbf{K}^{\mathrm{in}}$ and
$\mathbf{K}^{\mathrm{out}}$ are the wave vectors of the incident and
scattered waves, respectively) and $\mathbf{q}=\mathbf{Q}-\mathbf{Q}_0$
is a small deviation of $\mathbf{Q}$ from the nearest reciprocal lattice
vector $\mathbf{Q}_0$, so that $q\ll Q$. $\mathbf{U}(\mathbf{r})$ is the
sum of displacements produced by all defects of the crystal in a given
point $\mathbf{r}$ and $\left\langle \ldots \right\rangle $ denotes the
average over the statistics of the defect distribution. Equation
(\ref{eq3}) implies an infinite and statistically uniform sample, so that
the choice of origin is arbitrary. When finite size effects are
essential, as for example for misfit dislocations in epitaxial
layers,\cite{kaganer97} the correlation function $G(\mathbf{r}_{1},
\mathbf{r}_{2})$ depends on the difference of displacements $\mathbf{U}
(\mathbf{r}_{1}) - \mathbf{U}(\mathbf{r}_{2})$ and Eq.\ (\ref{eq2})
contains the exponent $\exp [i\mathbf{q}\cdot
(\mathbf{r}_{1}-\mathbf{r}_{2})]$.

We restrict ourselves to parallel straight dislocations and take
the direction of the dislocation lines as $z$ axis. Then,
$\mathbf{r}=(x,y)$ is a two-dimensional vector in the plane perpendicular
to the dislocation lines and Eq.\ (\ref{eq2}) can be written as
\begin{equation}
I(\mathbf{q})=\delta (q_{z})\int G(x,y)\exp (iq_{x}x+iq_{y}y)dxdy,
\label{eq2a}
\end{equation}
where the delta-function $\delta (q_{z})$ is due to the translational
invariance in the direction of the dislocation lines.

In the experiments described in the subsequent sections, the x-ray
scattering measurements from oriented samples are performed with a wide
open detector. The intensity (\ref{eq2}) is then to be integrated over
all directions of the scattered wave $\mathbf{K}^{\mathrm{out}}$. The
result of this integration is very similar to the powder average. The
actual part of the sphere $\left| \mathbf{K}^{\mathrm{out}}\right| =k$
(where $k$ is the wave vector) can be replaced by the plane perpendicular
to the direction of $\mathbf{K}^{\mathrm{out}}$. Integration of the
intensity (\ref{eq2a}) over this plane gives rise to a one-dimensional
integral
\begin{equation}
\mathcal{I}(q)=\int G(x)\exp (iqx/\cos \Phi )dx,  \label{eq7}
\end{equation}
where $\Phi $ is the angle between the $(x,y)$ plane and $\mathbf{K}^
{\mathrm{out}}$ (see Fig.\ \ref{skew}). It is given by $\sin \Phi =\sin
\Psi \sin \theta _{\mathrm{B}}$, where $\Psi $ is the angle between the
$(x,y)$ plane and the scattering vector $\mathbf{Q}$ and $\theta
_{\mathrm{B}}$ is the Bragg angle. The $x$ axis is chosen along the
projection of $\mathbf{K}^ {\mathrm{out}}$ on the plane perpendicular to
the dislocation lines. The corresponding expression for the case of
powder diffraction differs only by the direction of $x$: it runs along
$\mathbf{K}^{\mathrm{out}}$ for oriented films and along $\mathbf{Q}$ for
powder diffraction. The wave vector $q$ in Eq.\ (\ref{eq7}) is the
projection of $\mathbf{q}$ on the direction of
$\mathbf{K}^{\mathrm{out}}$, so that $q=Q \omega \cos \theta _{B}$, where
$\omega $ is the angular deviation from the peak center.

Krivoglaz\cite{krivoglaz:book,krivoglaz61,krivoglaz63} performed the
average over an uncorrelated defect distribution and showed that the
correlation function can be represented as
\begin{equation}
G(\mathbf{r})=\exp \left\{ -\rho \int \left[ 1-e^{i\mathbf{Q}\cdot
[\mathbf{u} (\mathbf{R+r})-\mathbf{u}(\mathbf{R})]}\right]
d\mathbf{R}\right\} .
\label{eq4}
\end{equation}
Here $\mathbf{u}(\mathbf{r})$ is the displacement field produced at a
point $\mathbf{r}$ by a single defect located at the origin and $\rho $ is
the defect density. For straight dislocations, the integration in Eq.\
(\ref{eq4}) is performed over the $\mathbf{R}=(X,Y)$ plane perpendicular
to the dislocation lines and $\rho $ is the dislocation density per unit
area. Thus, according to Eq.\ (\ref{eq4}), in order to obtain positional
correlations between two points separated by a distance $\mathbf{r}$, one
has to place a dislocation in an arbitrary position $\mathbf{R}$ and
perform the integration over all $\mathbf{R}$.

As a result of the slow decay of the dislocation strain, the main
contribution to the integral is due to remote dislocations, $R\gg r$, so
that the difference of displacements can be expanded as Taylor series,
$\mathbf{Q}\cdot [\mathbf{u}(\mathbf{R+r})-\mathbf{u}(\mathbf{R})]\approx
(\mathbf{r}\cdot \nabla )[\mathbf{Q}\cdot \mathbf{u}(\mathbf{R})]
=r_{i}Q_{j}\partial u_{j}/\partial X_{i}$. The distortion field of a
dislocation has a universal $r$-dependence, $\partial u_{j}/\partial
X_{i}=b\psi _{ij}/2\pi R$, where $b$ is the length of the Burgers vector
and $\psi _{ij}$ is a dimensionless factor of the order of unity which
depends only on the azimuth $\phi $. If the densities of dislocations
with opposite signs of the Burgers vectors are equal (there is no plastic
bend), the imaginary part of the correlation function (\ref{eq4}) is zero
and its real part, after expanding the exponent in curly brackets,
becomes\cite{krivoglaz63}
\begin{equation}
G(\mathbf{r})=\exp \left\{ -C\rho r^{2}\int \frac{dR}{R}\right\} ,
\label{eq5}
\end{equation}
where
\begin{equation}
C=\gamma (Qb)^{2}/4\pi ,\qquad \gamma =\frac{1}{2\pi }\int_{0}^{2\pi }
\left(\hat{r}_{i}\psi _{ij}\hat{Q}_{j}\right) ^{2}d\phi .
\label{eq6}
\end{equation}
$C$ and $\gamma $ are dimensionless factors of the order of unity. Here
$\mathbf{\hat{r}}$ and $\mathbf{\hat{Q}}$ are unit vectors in the
directions of $\mathbf{r}$ and $\mathbf{Q}$, respectively. The integral
in Eq.\ (\ref{eq5}) is taken from $\xi r$ (where $\xi \sim 1$ is a
dimensionless factor) to a limiting size $L$, which, for completely
uncorrelated dislocations, is equal to the crystal size. Then, the
integral is equal to $\ln (L/\xi r)$, and we arrive at Eq.\ (\ref{eq1}).
When the dislocations are correlated, so that the total Burgers vector
averaged over a certain characteristic scale is zero, the functional form
of the correlation function does not change but $L$ has the meaning of
that scale.\cite
{wilkens69,wilkens70nbs,wilkens70pss,krivoglaz83,krivoglaz:book}

The integration range in Eq.\ (\ref{eq7}) is limited by distances smaller
than $L$. A finite integration limit introduced in Eq.\ (\ref{eq7})
leads, in a numerical evaluation of the integral, to unphysical
oscillations in $\mathcal{I}(q)$ commonly appearing in Fourier integrals
taken over a rigidly limited interval. We found that an appropriate
smoothing is obtained by substituting $\ln (L/\xi x)$ in Eq.\ (\ref{eq1})
with $\ln [(L+\xi x)/\xi x]$ and extending the integration range to
infinity. The expressions for $\xi =\xi (\mathbf{Q})$ are bulky and
depend on the type of correlations in dislocation positions.\cite
{wilkens70nbs,wilkens70pss,krivoglaz:book,krivoglaz83} We restrict
ourselves to the first approximation that does not depend on the type of
correlations, $\xi =\left| \mathbf{Q}\cdot \mathbf{b}\right| /2\pi $.

Finally, combining the equations above, the diffracted intensity can be
represented as
\begin{eqnarray}
\mathcal{I}(\omega ) &=&I_{0}\int_{0}^{\infty }\exp (-Ax^{2}
\ln \frac{B+x}{x}) \cos (\omega x)dx  \nonumber \\
&+&I_{\mathrm{backgr}}.
\label{eq9}
\end{eqnarray}
We proceed here to the angular deviation from the peak maximum $\omega $
and introduce the peak height $I_{0}$ and the background intensity
$I_{\mathrm{backgr}}$ to provide the exact formula that is used for the
fits of the experimental peaks presented below. Parameters $A$
(describing the dislocation density) and $B$\ (describing the
dislocation correlation range) are given by
\begin{equation}
A=f\rho b^{2},\qquad B=gL/b.  \label{eq10}
\end{equation}
Here $f$ and $g$ are dimensionless quantities given by the diffraction
geometry,
\begin{equation}
f=\frac{\gamma }{4\pi }\frac{\cos ^{2}\Phi }{\cos ^{2}\theta _{B}},
\qquad g= \frac{2\pi \cos \theta _{\mathrm{B}}}{\cos \Phi \cos \Psi }.
\label{eq11}
\end{equation}
In Eq.~(\ref{eq11}), the expression for $g$ is written for edge
dislocations, taking into account the approximation $\xi =\left|
\mathbf{Q} \cdot \mathbf{b}\right| /2\pi $. For screw dislocations, $\cos
\Psi $ in this expression should be replaced by $\sin \Psi $. The length
of the Burgers vector $b$ in Eq.~(\ref{eq10}) is that of the relevant
Burgers vector for either edge or screw dislocations. The dimensionless
product $\rho b^{2}\ll 1$ is the mean number of dislocations crossing
each $b\times b$ cell in the plane perpendicular to the dislocation
lines. Equation (\ref{eq9}) with four parameters,
$A,B,I_{0},I_{\mathrm{backgr}}$ is used in Sec.\ \ref {sec:results}
below to fit the peak profiles and obtain the dislocation density $\rho $
and the length $L$.

\begin{figure}[tbp]
\noindent \includegraphics[width=\columnwidth]{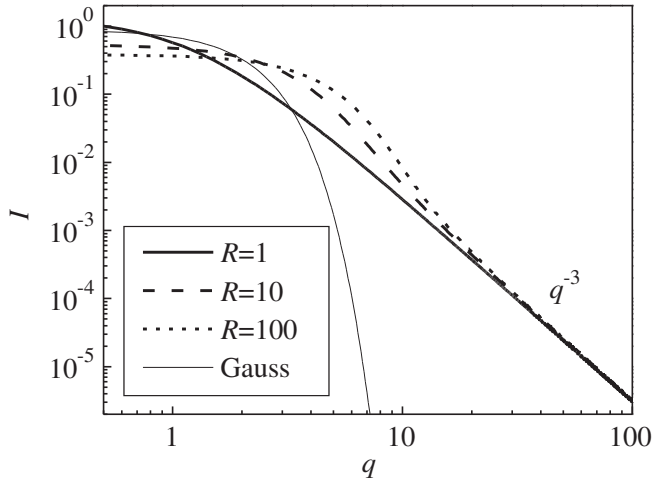}
\caption{Behavior of the integral (\ref{eq12}) for different values of
the parameter $R$. All curves merge at a common $q^{-3}$ asymptotic. A
Gaussian peak profile is shown by the thin line for comparison.}
\label{asymptotics}
\end{figure}

The behavior of the integral (\ref{eq9}) is illustrated in Fig.\ \ref
{asymptotics} where the function
\begin{equation}
I(q)=\int_{0}^{\infty }\exp \{-x^{2}\ln [(R+x)/x]\}\cos (qx)dx
\label{eq12}
\end{equation}
is numerically calculated for different values of the parameter $R$. The
curves merge at a common $q^{-3}$ asymptotic that does not depend on
$R$.\cite{krivoglaz:book,wilkens70pss,groma98,groma00} Then, Eq.\
(\ref{eq9}) has an asympotic behavior (for $\omega $ large in comparison
with the peak width)
\begin{equation}
\mathcal{I}(\omega )=A\frac{\pi I_{0}}{\omega ^{3}}+I_{\mathrm{backgr}}.
\label{eq13}
\end{equation}
We note that $\pi I_{0}$ is the integrated intensity of the peak.

Figure \ref{asymptotics} also shows that the asymptotics (\ref{eq13}) is
reached quite close to the peak center for $R\sim 1$, while for $R\gg 1$
the central part of the peak is Gaussian and the angular range where the
Gaussian approximation is valid increases with increasing $R$. The
FWHM of the calculated peaks increases with
increasing $R$ and can be approximated by
\begin{equation}
\Delta q\approx 2.4+\ln R.
\label{eq13a}
\end{equation}

Groma\cite{groma98} suggested to use the second restricted moment of the
intensity distribution
\begin{equation}
v_{2}(\omega )=\int_{-\omega }^{\omega }\varpi ^{2}[\mathcal{I}(\varpi )-
I_{\mathrm{backgr}}]d\varpi  \label{eq14}
\end{equation}
to obtain the dislocation density from the asymptotic behavior
(\ref{eq13}). Note that the integral (\ref{eq14}) diverges, when taken in
infinite limits. One finds, by substituting (\ref{eq13}) into
(\ref{eq14}), that
\begin{equation}
v_{2}(\omega )=2\pi I_{0}A\ln \omega +\mathrm{const.}
\label{eq15}
\end{equation}

It remains to calculate the orientational factor $C$ [Eq.\ (\ref{eq6})].
In the case of powder
diffraction,\cite{wilkens69,wilkens70nbs,wilkens70pss} the vector
$\mathbf{r}$ is directed along the projection $\mathbf{Q}_{\perp }$ of
the scattering vector $\mathbf{Q}$ on the $(x,y)$ plane. In our case, the
vector $\mathbf{r}$ is directed along the projection of
$\mathbf{K}^{\mathrm{out}}$ on the $(x,y)$ plane and makes an angle
$\alpha $ with the vector $\mathbf{Q}_{\perp }$, see Fig.\ \ref{skew}.
This angle is given by $\cos \alpha =\sin \theta _{\mathrm{B}}\cos \Psi
/\cos \Phi $. When evaluating the integral (\ref{eq6}) for edge
dislocations, we average $\gamma $ over possible orientations of the
Burgers vector in a hexagonal lattice (the dislocation lines are taken
along the sixfold axis) and obtain
\begin{equation}
\gamma _{\mathrm{e}}=\frac{9-8\nu +8\nu ^{2}-2(3-4\nu )\cos ^{2}\alpha }
{16(1-\nu )^{2}}\cos ^{2}\Psi ,  \label{eq8}
\end{equation}
where $\nu $ is the Poisson ratio. $\gamma _{\mathrm{e}}$ depends only
weakly on $\alpha $. Taking $\nu =1/3$, one can approximate $\gamma
_{\mathrm{e}}\approx \cos ^{2}\Psi $ in the whole range of reasonable
angles $\alpha $. The calculation for screw dislocations gives
\begin{equation}
\gamma _{\mathrm{s}}=\frac{1}{2}\sin ^{2}\Psi .  \label{eq16}
\end{equation}

Two limiting cases are of interest: a symmetric Bragg reflection to study
screw dislocations and a grazing incidence/grazing exit reflection as an
extreme case of a highly asymmetric skew geometry. For a symmetric Bragg
reflection, $\Psi =\pi /2$ and $\Phi =\theta _{B}$. Then, we obtain
$f=1/8\pi $ and, for screw dislocations, $g=2\pi $. The grazing
incidence/grazing exit geometry is the limit $\Phi =\Psi =0$, so that
$f=\gamma /(4\pi \cos ^{2}\theta _{B})$ and, for edge dislocations,
$g=2\pi \cos \theta _{\mathrm{B}}$.

The finite thickness $T$ of the epitaxial layer (with threading
dislocations perpendicular to the layer plane)\ can be taken into account
by making the Warren-Averbach ansatz\cite{WarrenAverbach50}
\begin{equation}
G(x)=G_{\mathrm{d}}(x)G_{\mathrm{s}}(x),  \label{eq17}
\end{equation}
where $G_{\mathrm{d}}(x)$ is the correlation function considered above.
The correlation function describing finite size effects can be written as
$G_{\mathrm{s}}(x)=\exp (-x/T)$. Its Fourier transformation is a
Lorentzian that is expected from the finite slit function $[\sin
(qT/2)/(qT/2)]^{2}$ after averaging over thickness variations. We note
that, in asymmetric reflections, Eq.\ (\ref{eq7}) gives rise to the
effective thickness $T/\cos \Phi $ along the direction of the diffracted
wave. The finite size correction (\ref{eq17}) does not complicate the
calculation of the intensity by numerical integration of Eq.\
(\ref{eq7}). If the resolution of the experiment cannot be neglected in
comparison with the peak width, the correlation function (\ref{eq17}) is
to be multiplied with the real-space resolution function
$\mathcal{R}(x)$, which also does not lead to additional complications of
the numerical integration.

\section{Experiment}

The GaN layers investigated here were grown on 6H-SiC(0001) by
plasma-assisted molecular beam epitaxy (PAMBE). The two PAMBE systems
employed are equipped with a solid-source effusion cell for Ga and a
radio-frequency nitrogen plasma source for producing active N. Both
systems have a base pressure of 5$\times $10$^{-11}$~Torr. We use 6N
N$_{2}$ gas as a precursor which is further purified to 5~ppb by a getter
filter. H$_2$-etched 6H-SiC(0001) wafers produced by Cree\texttrademark\
were used as substrates.\cite{brandt4019apl} An \emph{in situ} Ga
flash-off procedure was performed in order to remove residual suboxides
from the SiC substrate surface prior to growth.\cite{brandt4019apl} The
temperatures were calibrated by visual observation of the melting point
of Al (660$^{\circ }$C) attached to the substrate.

\begin{figure}[tbp]
\noindent \includegraphics[width=\columnwidth]{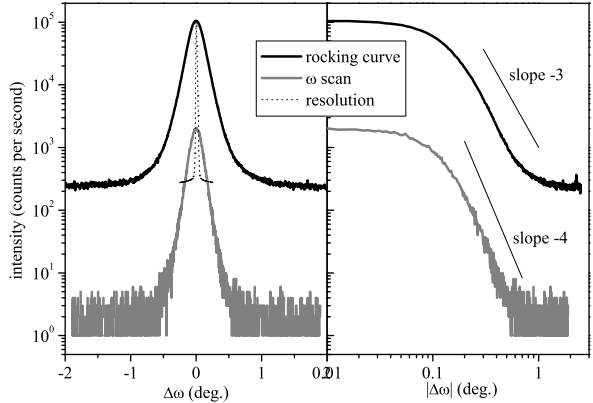}
\caption{Double-crystal rocking curve with open detector (black line) and
triple-crystal $\omega$ scan with analyzer (gray line) recorded in skew
geometry across the 10.5 reflection of sample 1. The symmetric 004
reflection from a perfect Si crystal (broken line) is shown as a measure
of the resolution. The intensities from the GaN layer are presented in
counts per second, the Si(004) signal is scaled appropriately. The right
panel shows the same peaks in log-log scale.}
\label{omegascan}
\end{figure}

Sample 1 was grown under Ga-stable conditions with a substrate
temperature of 740$^{\circ}$C. The growth rate employed was 140~nm/h, and
the film was grown to a final thickness of 340~nm. Under these growth
conditions, we observed an entirely streaky, (1$\times$1) RHEED pattern
throughout growth except for the initial nucleation stage. The surface
morphology of the film as observed by atomic force microscopy (AFM)
exhibits clearly resolved monatomic steps. The root-mean-square roughness
amounts to 0.3~nm over an area of $1 \times 1$~$\mu$m$^2$. Samples grown
under these conditions typically exhibit a narrow symmetric reflection,
suggesting a low density of screw dislocations. In contrast, sample 2 was
grown under near-stoichiometric conditions with a substrate temperature
of 780$^{\circ}$C. The growth rate employed was 445~nm/h and the total
thickness of the film amounts to 1660~nm. After the initial 100~nm of GaN
growth, a 10~nm thick AlN film was deposited prior to overgrowth by GaN.
The RHEED pattern exhibited a superposition of streaks and $\vee$-shaped
chevrons indicative of facetting. Indeed, the AFM micrographs of this
sample showed the characteristic plateau-valley morphology of GaN films
grown with insufficient Ga flux. Compared to films grown under the same
conditions as sample 1, films grown under these conditions exhibit
narrower asymmetric reflections, indicating a reduction of the edge
dislocation density.

X-ray measurements were performed with a Philips X'Pert
PRO\texttrademark\ four-circle triple-axis diffractometer equipped with a
Cu$K_{\alpha 1}$ source in the focus of a multilayer x-ray mirror and a
Ge(022) hybrid monochromator. The detector was kept wide open and at
fixed position $2\theta_B$, leading to an angular acceptance of $1^{\circ
}$. All asymmetric rocking curves were recorded in skew
geometry.\cite{sun02}

We denote the GaN reflections in the form \textit{hk.l} which is
equivalent to the four-index notation \textit{hkil} for hexagonal
crystals with $h+k+i=0 $.

\section{Results}
\label{sec:results}

\subsection{X-ray diffraction}

\begin{figure}[tbp]
\noindent \includegraphics[width=\columnwidth]{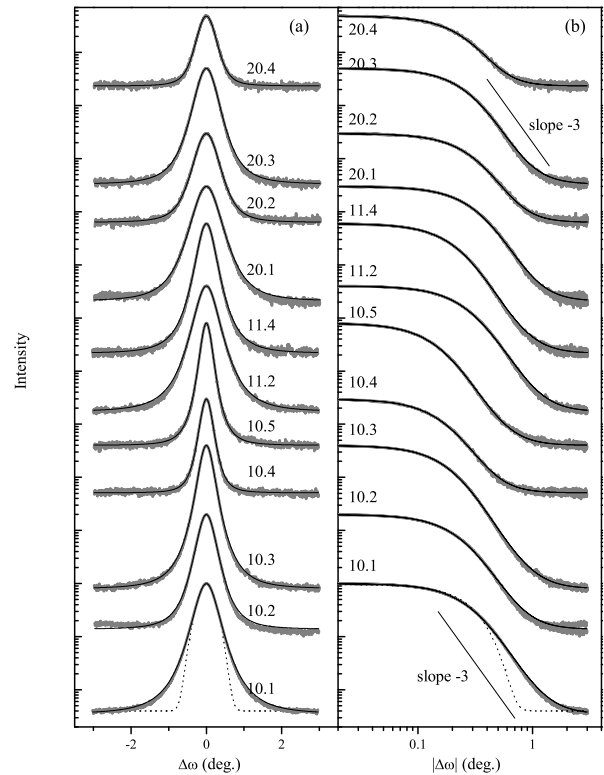}
\caption{Double-crystal rocking curves from sample 1 obtained in skew
geometry for various different reflections as indicated in the figure.
The profiles are shown in logarithmic (a) and log-log (b) scales. The
experimental data are shown by gray lines. The full black lines are fits
of the intensity by Eq.\ (\ref{eq9}). The dotted lines for the 10.1
profile show a Gaussian profile.}
\label{asymm_peaks}
\end{figure}

Figure \ref{omegascan} compares a double-crystal rocking curve measured
with open detector (acceptance 1$^{\circ }$) and a triple-crystal
$\omega$ scan measured with a three-bounce Ge(022) analyzer across the
10.5 reflection for the same GaN film (sample 1). The comparison with the
004 rocking curve from a perfect Si crystal shows that the GaN peak
broadening due to instrumental resolution can be neglected. The log-log
plots (right column) show that the tails of the intensity distributions
follow the asymptotic laws expected for scattering from dislocations,
$I\propto q^{-3}$ for measurements with the open detector and $I\propto
q^{-4}$ in the case of a collimated diffracted beam. Thus, the comparison
of these two profiles shows that they contain the same information about
lattice distortions in the film. The rocking curve measurements with open
detector have, however, both experimental and theoretical advantages. The
experimental advantage is a two orders of magnitude higher intensity (the
intensities are plotted in Fig.\ \ref{omegascan} in counts per second).
The analysis of the rocking curve is also more simple since the intensity
distribution is described by the one-dimensional integral (\ref{eq7}),
while the analysis of the scans with the analyzer requires the
two-dimensional integration (\ref {eq2a}) of the correlation function.
Therefore, we restrict ourselves to the analysis of double-crystal
rocking curves.

\begin{figure}[tbp]
\noindent \includegraphics[width=\columnwidth]{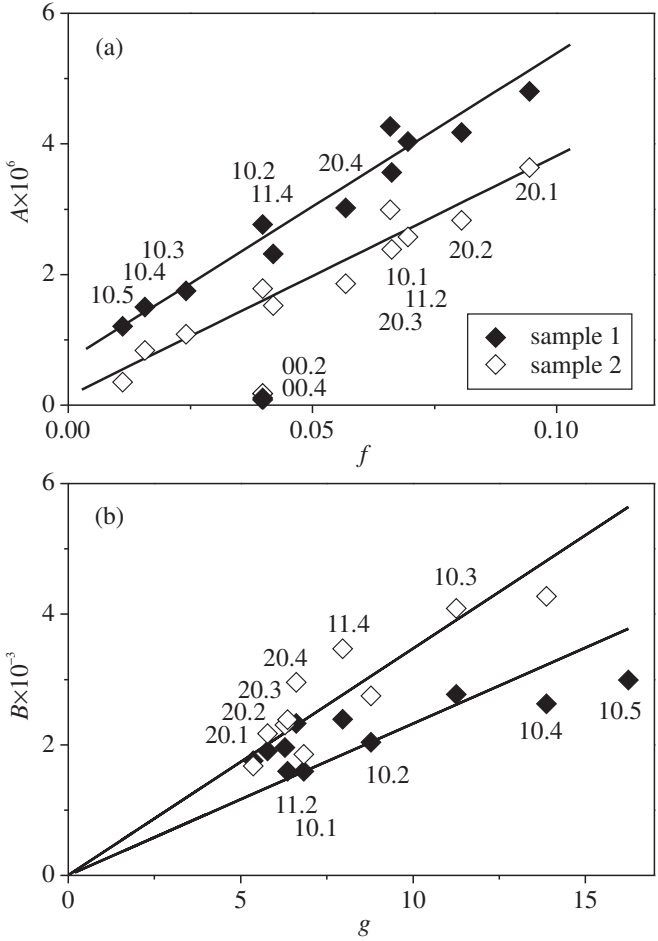}
\caption{Parameters $A$ and $B$ obtained from the fits of the
experimental profiles by Eq.\ (\ref{eq9}). Samples 1 and 2 are denoted by
full and open symbols, respectively.}
\label{parametersAB}
\end{figure}

Figure \ref{asymm_peaks} presents skew-geometry rocking curves for
various reflections from sample 1. Several conclusions can be drawn just
from a direct inspection of the peak profiles. First, the lineshapes are
far from being Gaussian at the tails of the intensity distribution: see
the comparison of the 10.1 profile with a Gaussian fit [dotted lines in
Figs.\ \ref {asymm_peaks} (a,b)]. When the dynamic range is larger than
two orders of magnitude, it is evident that the tails of the asymmetric
profiles rather follow the $q^{-3}$ law. For relatively weak reflections
(e.~g., 10.4 or 20.4), the $-3$ exponent is not reached. Thus, the
profiles obey the behavior typical for a random dislocation distribution.
In the following, we analyze them quantitatively to obtain the
characteristics of the dislocation ensemble.

\begin{figure}[tbp]
\noindent \includegraphics[width=\columnwidth]{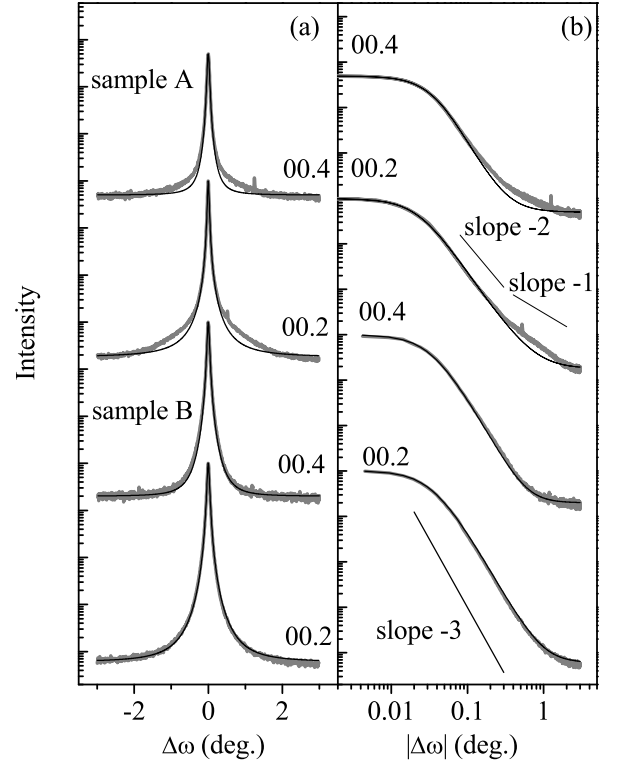}
\caption{Symmetric 00.2 and 00.4 x-ray diffraction profiles from samples
1 and 2. The profiles are shown in logarithmic (a) and log-log (b) scales.}
\label{symm_peaks}
\end{figure}

The solid lines in Figs.\ \ref{asymm_peaks} (a) and (b) are the fits of
the measured profiles by Eq.\ (\ref{eq9}). One can see that the peak
profiles are adequately described. In Fig.\ \ref{parametersAB}, we plot
the fit parameters $A$ and $B$ as functions of $f$ and $g$, respectively,
since according to Eq.\ (\ref{eq10}) both dependencies are expected to be
linear. Figure \ref{parametersAB} (a) can be considered as a refined
version of the Williamson-Hall plot. The linear fit of the data in Fig.\
\ref{parametersAB} for sample 1 crosses the axis of the ordinates at a
small but non-zero value of $A$, which indicates that, in addition to
threading dislocations, there is an additional source of peak broadening,
namely, size broadening.  This effect is much smaller for sample 2. From
the slopes of the straight lines in Fig.\ \ref{parametersAB} (a), we
obtain $\rho _{\mathrm{e}}b_{\mathrm{e}}^{2}=A/f=4.7\times 10^{-5}$ and
$3.6\times 10^{-5} $ for samples 1 and 2, respectively, where
$b_{\mathrm{e}}=0.32$~nm is the length of the Burgers vector of edge
dislocations. This result yields a density of edge threading dislocations
$\rho _{\mathrm{e}}=4.6\times 10^{10}$ cm$^{-2}$ for sample 1 and
$3.5\times 10^{10}$ cm$^{-2}$ for sample 2. The mean distances between
edge dislocations are $r_{d}=1/\sqrt{\rho _{\mathrm{e}}}=47$~nm (sample
1) and 53~nm (sample 2). From the slopes of the lines in Fig.\
\ref{parametersAB} (b), we obtain $L/b_{\mathrm{e}}=B/g=230$ and $330$,
which give the characteristic lengths of the dislocation correlations
$L=74$~nm and $106$~nm, respectively. The dimensionless parameter
characterizing the dislocation correlations\cite{wilkens70pss} possesses
the values $M=L/r_{d}=1.6$ for sample 1 and 2.0 for sample 2.

We also fitted the profiles with Eqs.~(\ref{eq13}) and (\ref{eq15}), which
require less computational efforts. These fits give slightly larger
values for parameter $A$ (dislocation density) and are not able to give
parameter $B$ for dislocation correlations.

\begin{figure}[tbp]
\noindent \includegraphics[width=\columnwidth]{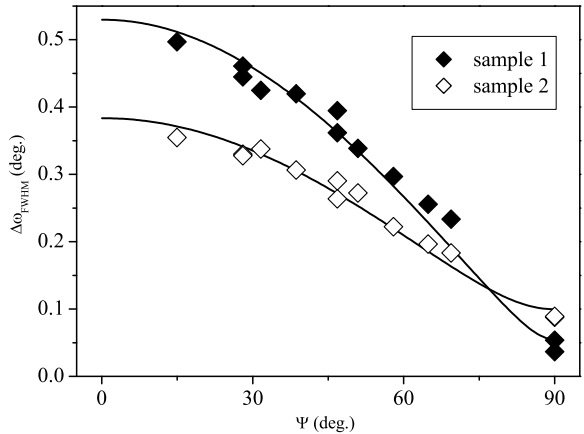}
\caption{FWHM of the rocking curves as a function of inclination angle
$\Psi$ for the two GaN films. The symbols are experimental data and the
lines are fits to them.}
\label{twist}
\end{figure}

The symmetric x-ray diffraction profiles shown in Fig.\ \ref{symm_peaks}
are narrow compared to the asymmetric reflections. Furthermore, the
profiles of samples 1 and 2 are qualitatively different. In sample 1, the
intensity distributions obey a $q^{-2}$ law in the intermediate range of
angular deviations (and intensities) that is followed by an even lower
exponent for large deviations (and very low intensities). In sample 2,
the intensity distributions are close to the $q^{-3}$ law.

Edge threading dislocations with dislocation lines normal to the surface
and Burgers vectors in the surface plane, which are the main source of
the peak broadening in asymmetric reflections, do not distort the planes
parallel to the surface. They do thus not contribute to diffraction in
the symmetric reflections. The $q^{-2}$ law for the 00.2 and 00.4 peaks
of sample 1 points to the finite thickness of the epitaxial layer as the
main source of the broadening. The $q^{-1}$ law at the far tails of the
peaks are due to another source, possibly thermal diffuse scattering. We
did not investigate this part of the symmetric reflections since we
suppose that it is not related to threading dislocations which are the
topic of the present study. The profiles of sample 2, which is three
times thicker than sample 1, obey the $q^{-3}$ law indicating that the
broadening is primarily by dislocations.

The solid lines in Fig.\ \ref{symm_peaks} are the fits of the
experimental curves to Eq.\ (\ref{eq17}), where the different thicknesses
of the layers are explicitly taken into account. We indeed obtain the
$q^{-2}$ intensity decay on the tails of the peak for sample 1 and the
$q^{-3}$ decay for sample 2. From the fit parameters, we obtain  $\rho
_{\mathrm{s}}b_ {\mathrm{s}}^{2}=A/f=2.4\times 10^{-6}$ and $4.5\times
10^{-6}$ for samples 1 and 2, respectively, where
$b_{\mathrm{s}}=0.52~$nm is the length of the Burgers vector of screw
dislocations. This results yields the screw threading dislocations
densities of $\rho _{\mathrm{s}}=9\times 10^{8}$ cm$^{-2}$ for sample 1
and $1.7\times 10^{9}$ cm$^{-2}$ for sample 2. From the values of the
parameter $B$ we obtain $L/b_{\mathrm{s}}=B/g=650$ and $440$, which
result in characteristic lengths of the dislocation correlations
$L=340$~nm and $230$~nm for sample 1 and 2, respectively. The parameter
$M$ is close to 1 for both samples. Note, however, that screw
dislocations are not the only source of the peak broadening in symmetric
reflections (see discussion below in Sec. \ref{sec:discuss}).

\subsection{TEM}

TEM is the method of choice to directly determine the character of
dislocations and their distribution in thin films. The $\mathbf{g \cdot
b}$ criterion in TEM is generally applied for the two-beam condition to
evaluate the alignment of the strain field of the dislocation with
Burgers vector $\mathbf{b}$ with respect to the diffraction vector
$\mathbf{g}$ producing the image contrast. This criterion is strictly
correct for screw dislocations and for edge dislocations only if their
line direction $\mathbf{l}$ and Burgers vector are in the imaging plane,
i.~e., if $\mathbf{g \cdot (b \times l)}$ is considered. In the present
case of GaN(0001) films it is well established that three types of
threading dislocations exist, having Burgers vectors $\frac{1}{3}
\left\langle 11\overline{2}0\right\rangle$, $\left\langle 0 0 0
1\right\rangle$ and $\frac{1}{3} \left\langle
11\overline{2}3\right\rangle$ representing edge, screw and mixed
dislocations, respectively, under the assumption that the dislocations
lines lie parallel to the $c$-axis. Mixed dislocations are not observed in
our samples.

\begin{figure}[t!]
\noindent \includegraphics[width=\columnwidth]{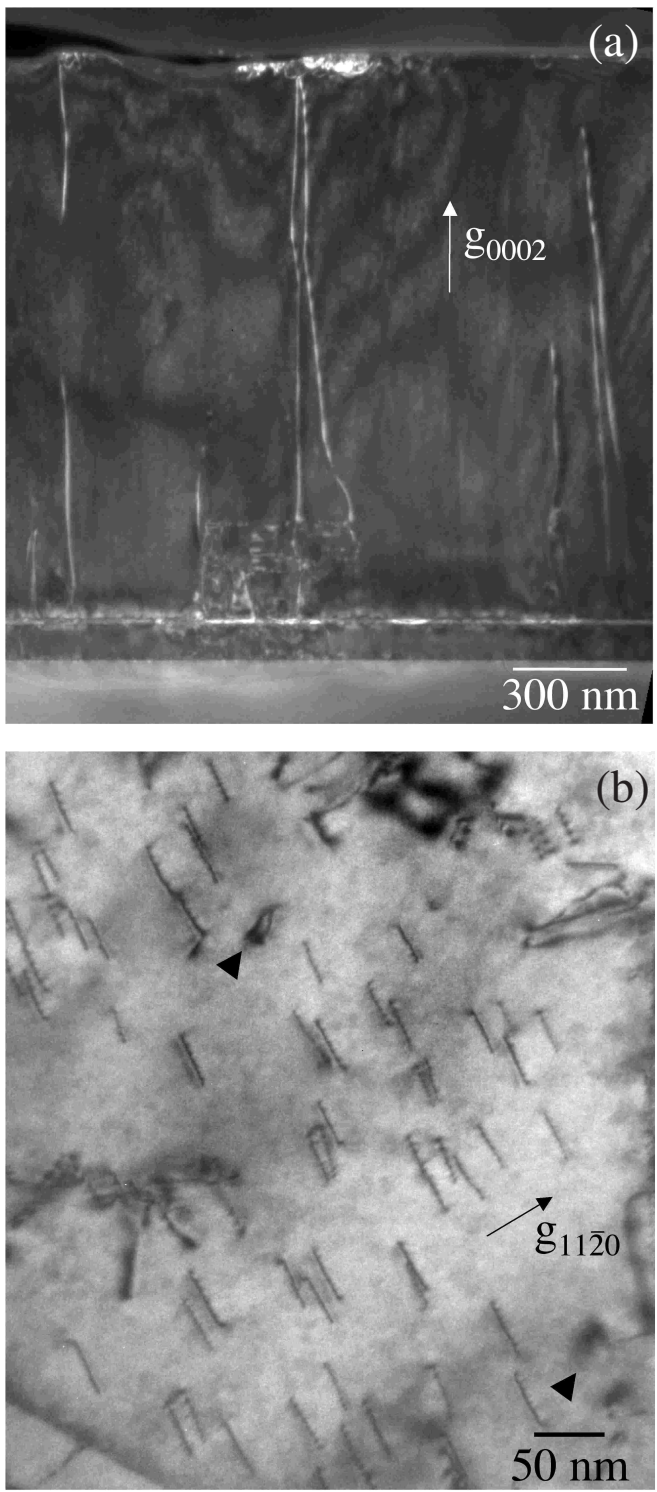}
\caption{Cross-sectional (a) and plan-view (b) TEM images of sample 2.
The outcrops of screw dislocations are marked in (b) by arrows.}
\label{TEM}
\end{figure}

In order to identify the Burgers vector, two images have to be recorded
with $\mathbf{g}$ parallel and perpendicular to the $c$-axis. The screw
dislocations are thus imaged if $\mathbf{g}$=[0002], an example of which
is shown in Fig.\ \ref{TEM}(a) for sample 2. From this image, we can
directly measure the dislocation density if we know the TEM specimen
thickness that is determined by tilting the interface from the end-on to
a well-defined inclined position.  Edge dislocations appear in the
cross-sectional images if $\mathbf{g}$ is perpendicular to the $c$-axis,
e.~g., if $\mathbf{g}=[11\overline{2}0]$, and we are then able to measure
their density in the same way. Furthermore, plan-view TEM imaging is
applied to complement the measurement of the dislocation density. Figure
\ref{TEM}(b) shows a plan-view TEM image of sample 2, where the specimen
is tilted a few degrees off the [0001] zone axis to obtain two-beam
conditions with $\mathbf{g}=[11\overline{2}0]$ in order to bring the edge
dislocations in contrast. The edge threading dislocations are extended
along the film normal and homogeneously distributed. It is remarkable
that screw dislocations can be identified as well (marked by arrows),
although the contrast should vanish because of $\mathbf{g \cdot b}=0$.
However, this specific contrast is produced due to strain relaxation of
the screw dislocation at the free surfaces varying locally the lattice
plane distortions created by the strain field. Table \ref{table} compares
the densities of edge and screw dislocations obtained from the x-ray
diffraction and TEM measurements, revealing that the density of edge
dislocations is in fact slightly higher in sample 1 compared to sample 2.
The screw dislocation density in sample 1 is too low for reliable TEM
determination. Some of the screw dislocations are seen as hexagonal pits
in AFM micrographs. A lower limit of 5$\times$10$^7$~cm$^{-2}$ is thus
obtained for the screw dislocation density in sample 1.

\begin{table}[tb]
\caption{Edge and screw dislocation densities determined by x-ray
diffraction and TEM.}
\label{table}
\begin{ruledtabular}
\begin{tabular}{lcllll}
&thickness& \multicolumn{2}{c}{$\rho _{\mathrm{e}}$ ($10^{10}$ cm$^{-2}$)}
  &\multicolumn{2}{c}{$\rho _{\mathrm{s}}$ ($10^{ 9}$ cm$^{-2}$)}   \\
&(nm)& X-ray & TEM & X-ray & TEM \\
\hline
sample 1 & 340 & 4.6 & 3.0$\pm$0.5 & 0.9 & --- \\
sample 2 & 1660 & 3.5 & 2.0$\pm$0.5 & 1.7 & 1.2 \\
\end{tabular}
\end{ruledtabular}
\end{table}

\section{Discussion}

\label{sec:discuss}

The x-ray diffraction studies of GaN have used, up to now, only the
widths of diffraction peaks. Two quantities can be obtained from a series
of the reflections.\cite{sun02} One is the FWHM of the symmetric
reflection $\Delta \omega _{\mathrm{s}}$ which is influenced by screw
dislocations but insensitive to edge dislocations. The other quantity is
obtained by extrapolation of the FWHM of skew reflections to the limiting
case of grazing incidence/grazing exit diffraction, with both incident
and diffracted beams lying in the surface plane. This quantity ($\Delta
\omega _{\mathrm{e}}$) is sensitive to edge but insensitive to screw
dislocations. Figure \ref{twist} presents the widths of the diffraction
profiles of samples 1 and 2 as a function of the inclination angle
$\Psi$. The lines are fits by the model described in Ref.\
\onlinecite{sun02} to obtain $\Delta \omega _{\mathrm{s}}$ and $\Delta
\omega _{\mathrm{e}}$. While these quantities are determined
reliably and accurately, the question arises how they are related to the
actual dislocation densities. Most
commonly,\cite{metzger1013pma,chierchia03} the formulas initially
proposed in Refs.\ \onlinecite{gay53} and \onlinecite{dunn57} for a mosaic
crystal are used for this purpose:
\begin{equation}
\rho _{\mathrm{e}}=\frac{\Delta \omega _{\mathrm{e}}^{2}}{4.35b_
{\mathrm{e}}^{2}},\qquad \rho _{\mathrm{s}}=\frac{\Delta
\omega _{\mathrm{s}}^{2}} {4.35b_{\mathrm{s}}^{2}},
\label{eq18}
\end{equation}
with the coefficient $2\pi \ln 2\approx 4.35.$ In this context, the
quantities $\Delta \omega _{\mathrm{s}}$ and $\Delta \omega
_{\mathrm{e}}$ are often referred to as tilt and twist, respectively.

Equations (\ref{eq18}) are valid for mosaic crystals, where the terms
tilt and twist describe different modes of relative block misorientation.
These terms loose their meaning when applied to a layer with randomly
distributed dislocations, since the broadening is then determined by a
complicated combination of misorientation of the lattice planes and
change of interplanar distances (strain). Hence, equations (\ref{eq18})
use an inappropriate approach and are not strictly valid in the case of a
random dislocation distribution. It remains correct that the dislocation
density is proportional to the square of the peak width and inversely
proportional to the square of the relevant Burgers vector, but the
coefficients need to be reconsidered.

The relation between the dislocation density and the FWHM $\Delta \omega $
of the intensity distribution (\ref{eq9}) can be written, by making use of
Eqs.\ (\ref{eq10}) and (\ref{eq13a}), in a form similar to Eq. (\ref{eq18}):
\begin{equation}
\rho \approx \frac{\Delta \omega ^{2}}{[2.4+\ln (g\sqrt{f}M)]^{2}fb^{2}}.
\label{eq19}
\end{equation}
The dimensionless parameter $M=L\sqrt{\rho }$ that characterizes
dislocation correlations is slightly larger than 1 for the GaN films
studied here, which is an indication of a strong screening of the
long-range strain fields of dislocations by the neighboring dislocations.
One has $M\gg 1$ for uncorrelated dislocations. In this latter case, the
logarithmic term in Eq.\ (\ref{eq19}) is the one obtained by
Krivoglaz.\cite {krivoglaz63,krivoglaz:book}

We can now simplify Eq. (\ref{eq19}) for the limiting cases of the
large-inclination skew diffraction and for symmetric Bragg diffraction by
using the expressions for the parameters $f$ and $g$ obtained in Sec.\
\ref {sec:theory} for these cases:
\begin{equation}
\rho _{\mathrm{e}}\approx \frac{4 \pi \Delta \omega _{\mathrm{e}}^{2}\cos
^{2}\theta _{B}}{(3.0+\ln M)^2 b_{\mathrm{e}}^{2}},
\qquad \rho _{\mathrm{s}}\approx
\frac{8 \pi \Delta \omega _{\mathrm{s}}^{2}}{(2.6+\ln M)^2
b_{\mathrm{s}}^{2}}.
\label{eq20}
\end{equation}
The term $\ln M$ describing the range of dislocation correlations cannot
be neglected even when $M$ varies between 1 and 2, as it happens to be
the case for the samples studied in the present work. Compared to Eqs.\
(\ref{eq18}), Eqs.\ (\ref{eq20}) result in a four times higher edge
dislocation density and an order of magnitude higher screw dislocation
density and give dislocation densities that are in agreement with the TEM
results, see Table \ref{table}.

There are at least three types of dislocations that may contribute to the
broadening of symmetric reflections. First, screw threading dislocations
distort the lattice planes parallel to the surface. Secondly, edge
threading dislocations can contribute to these reflections if the
dislocation lines deviate from the layer normal. Even if such deviations
are small, the effect cannot be neglected since the density of edge
threading dislocations is much larger than the density of screw threading
dislocations. Third, misfit dislocations at the layer-substrate interface
contribute to broadening of the symmetric reflections. Although misfit
dislocation lines are far from the surface, their non-uniform distortions
penetrate through the whole epitaxial layer.\cite {kaganer97} Finally,
the stress relaxation at the free surface gives rise to additional
strains around the outcrops of the edge threading dislocations,
contributing to an additional broadening to the symmetric reflections.
Thus, Eq.~(\ref{eq9}) and, particularly, Eqs.~(\ref{eq20}) provide an
upper estimate of the screw threading dislocation density since they are
assumed to be the only source of broadening for symmetric reflections.

\section{Conclusions}

The width of either symmetric or asymmetric reflection can be used as a
figure-of-merit for the dislocation density only if the dislocation
distribution is the same in all the samples to be compared. Even for
films having a spatially random distribution of dislocations, the width
of a given reflection depends not only on the dislocation density, but
also on the range of correlations in the restricted random dislocation
distribution. The lineshape analysis of the diffraction profile as
presented in this work returns the width as well as the correlation
range, and is thus a far more reliable approach for estimating the
dislocation density than a simple consideration of the width alone.

The lineshape analysis has shown that the x-ray diffraction profiles of
the GaN films under investigation are Gaussian only in the central part
of the peak. The tails of the peak follow the power laws characteristic
to x-ray diffraction of crystals with randomly distributed dislocations.
The double-crystal rocking curves measured with a wide open detector
follow a $q^{-3}$ behavior, while the triple-crystal rocking curves with
an analyzer crystal obey a $q^{-4}$ behavior. The study of the
double-crystal rocking curves is more simple both experimentally and
theoretically, since the diffracted intensity is larger and the peak
profile is described by a one-dimensional Fourier transform of the pair
correlation function (\ref{eq9}).

The $q^{-3}$ tails of the diffraction profiles are insensitive to
correlations between dislocations and allow a more reliable determination
of the dislocation densities. The entire diffraction profiles are
adequately fitted by Eq.\ (\ref{eq9}). The fits provide two parameters
characterizing the dislocation ensemble, the mean dislocation density
$\rho $ and the screening range $L$. The latter quantity corresponds to a
mean size of the cells with the total Burgers vector equal to zero. We
find that, for edge threading dislocations in GaN layers, $L$ is only
slightly larger than the mean distance between dislocations $\rho
^{-1/2}$.

%\bibliographystyle{apsrev}
%\bibliography{surface,twist}

\end{document}